\documentclass[sigchi]{acmart} %review,

\usepackage{booktabs} % For formal tables

\usepackage{array,multirow}

\usepackage{boldline}
\usepackage{graphicx}
\usepackage{xcolor}

\graphicspath{{./figs/}}

\setcopyright{none}
\AtBeginDocument{%
  \providecommand\BibTeX{{%
    \normalfont B\kern-0.5em{\scshape i\kern-0.25em b}\kern-0.8em\TeX}}}

\begin{document}
\title{Understanding Relations Between Perception of Fairness and Trust in Algorithmic Decision Making}
\renewcommand{\shorttitle}{Understanding Relations Between Perception of Fairness and Trust}
\titlenote{This paper has been accepted by The International Conference on Behavioral and Social Computing (BESC 2021), October 2021}
% \subtitle{Extended Abstract}
% \subtitlenote{The full version of the author's guide is available as
%   \texttt{acmart.pdf} document}

\author{Jianlong Zhou, Sunny Verma, Mudit Mittal, and Fang Chen}
\authornote{S. Verma and M. Mittal: Work done while working at UTS.}
\affiliation{%
  \institution{Data Science Institute, University of Technology Sydney}
%   \streetaddress{13 Garden Street}
  \city{Sydney}
  \state{NSW}
  \country{Australia}
  \postcode{2007}
}
\email{Jianlong.Zhou@uts.edu.au}

% \author{Fang Chen}
% \affiliation{%
%   \institution{DATA61, CSIRO}
%   \streetaddress{13 Garden Street}
%   \city{Eveleigh}
%   \state{NSW}
%   \country{Australia}
%   \postcode{2015}
% }
% \email{fang.chen@data61.csiro.au}

\renewcommand{\shortauthors}{Zhou et al.}

\begin{abstract}
Algorithmic processes are increasingly employed to perform managerial decision making, especially after the tremendous success in Artificial Intelligence (AI). This paradigm shift is occurring because these sophisticated AI techniques are guaranteeing the optimality of performance metrics. However, this adoption is currently under scrutiny due to various concerns such as fairness, and how does the fairness of an AI algorithm affects user's trust is much legitimate to pursue. In this regard, we aim to understand the relationship between induced algorithmic fairness and its perception in humans. In particular, we are interested in whether these two are positively correlated and reflect substantive fairness. Furthermore, we also study how does induced algorithmic fairness affects user trust in algorithmic decision making. To understand this, we perform a user study to simulate candidate shortlisting by introduced (manipulating mathematical) fairness in a human resource recruitment setting. Our experimental results demonstrate that different levels of introduced fairness are positively related to human perception of fairness, and simultaneously it is also positively related to user trust in algorithmic decision making. Interestingly, we also found that users are more sensitive to the higher levels of introduced fairness than the lower levels of introduced fairness. Besides, we summarize the theoretical and practical implications of this research with a discussion on perception of fairness.

% Artificial Intelligence (AI) approaches such as machine learning algorithms are increasingly used to make managerial decisions that people used to make. Perception of AI characteristics such as fairness can significantly influence adoptions of AI solutions, yet we do not fully understand how AI fairness affects user trust. 
% This paper aims to understand whether the introduced fairness can be positively perceived by humans to reflect the actual fairness of AI models, and further investigates how the introduced fairness affects user trust in algorithmic decision making. 
% A user study on the candidate shortlisting in a human resource recruitment setting was conducted by manipulating different levels of introduced fairness. Experimental results found that introduced fairness was positively related to human perceived fairness, and the introduced fairness was positively related to user trust in algorithmic decision making. 
% It was also found that users were more sensitive to the high level of introduced fairness than the low level of introduced fairness. Finally, the theoretical and practical implications of this research are discussed. 
\end{abstract}



%, algorithmic decision making}

%  \caption{This is a teaser}

\maketitle

\section{Introduction}

Artificial Intelligence (AI) has powerful capabilities in prediction, automation, planning, targeting, and personalisation \cite{zhou_ai_2019}.
It has been increasingly used to make important decisions that affect human lives in different areas ranging from social and public management to promoting productivity for economic wellbeing.
For example, AI can be used to decide the loan approval in banks and manage engagement and outcomes of job for workers within an organization. These algorithms are also utilized by various hiring platforms to recommend and recruit candidates in human resource settings
\cite{hughes_artificial_2019,gugnani_implicit_2020} (such AI-informed decision making is also called algorithmic decision making). 
Besides all these functionalities of AI, a paramount concern with AI's decision making is equal treatment or equitably of decision based on people’s performance or needs \cite{leventhal_what_1980} is required \cite{lavelle_fairness_2009,robert_designing_2020}. This setting of equitable treatment is also known as fairness in AI. On the other hand, unintentional (or intentional) discrimination can cause unfairness in AI and lead to poor decision making. Thus fairness becomes critical as a fair decision making system amplifies the satisfaction levels with algorithmic decision making \cite{lavelle_fairness_2009,robert_designing_2020}. Often, the fairness is a consequence of either the training data or the design of machine learning models, which is the fairness human actually perceives in algorithmic decision making, ultimately affect their adoptions in real-world applications \cite{lee_understanding_2018}.

% Fairness is defined as a global perception of appropriateness -- a perception that tends to lie theoretically downstream of justice \cite{colquitt_measuring_2015}.

Besides, inputs to AI models such as machine learning models are often historical records or samples of events. They are usually not the precise description of events and conceal discrimination with sparse details which are very difficult if not impossible to identify. AI models are also imperfect abstractions of reality as their sole purpose if better generalization capabilities. Therefore, the imprecision and uncertainty associated with AI are imminent. Meanwhile, AI models are usually ``black-boxes'' for users and even for AI experts \cite{Zhou_human_2018}. Users simply provide input data to an AI system, and after selecting some menu options, the system displays colourful viewgraphs and/or recommendations as output. It is neither clear nor well understood why these AI algorithms made a certain prediction, or how trustworthy their prediction are. In a nutshell, the concerns demonstrate that the successful use of AI critically depends on user trust in AI systems.
One of widely cited definition defines trust as ``the willingness of a party to be vulnerable to the actions of another party based on the expectation that the other will perform a particular action important to the trustor, irrespective of the ability to monitor or control that other party'' \cite{mayer_integrative_1995}. 
Considerable research on fairness has evidenced that fairness perceptions are linked to trust such as in management and organizations \cite{komodromos_employees_2014,roy_impact_2015}.

% Various research suggests that fairness influence people's trust as well as decision-making process \cite{konovsky_citizenship_1994,roy_impact_2015}.
% cohen-charash_role_2001,

Different from above, in algorithmic decision making, mathematical fairness introduced by AI models and/or data (also refers to \emph{introduced fairness} in this paper) is perceived by humans (also refers to \emph{perception of fairness} in this paper) implicitly or explicitly.
The perceived fairness is a central component of maintaining satisfactory relationships with humans in decision making \cite{aggarwal_when_2012}.
Given various mathematical formulations of fairness, three major findings are: 1) demographic parity most closely matches human perception of fairness \cite{srivastava_mathematical_2019}; 2) effects of transparency and outcome control on perceived fairness \cite{lee_procedural_2019}; and 3) factors affecting perceptions of fairness in algorithmic decision making \cite{wang_factors_2020}. While the fairness (or discrimination) is either introduced by AI models and/or the data, it is critical to understand whether an introduced level of fairness is affecting its perception by humans in algorithmic decision making. Therefore, in this work we aim to investigate the relations between the introduced fairness and human perception of fairness. 

% Further is introduced fairness level is equivalent to the actual fairness level (measured by some metric) of an AI algorithm?

Our aim in this paper is to understand what is the perception of introduced fairness by humans in particular, is it positive or negative? Importantly, we further dwell to understand whether the introduced fairness affects users trust in algorithmic decision making. In this regard, we utilise the statistical parity as the actual fairness level of an AI system (defined in sec.~\ref{sec:preliminary}, Eq.~\ref{Statsparity}) as it has been widely accepted as a metric to measure fairness. We then design a user study to investigate the perception of fairness by simulating a human resource recruitment for candidate shortlisting by manipulating introduced fairness. Due to lock-down restrictions, this user study was performed online to collect the participant responses during the COVID-19 pandemic.
In summary, our experimental results demonstrate that two important findings: 1) introduced fairness is positively related to human perception; and 2) simultaneously, high level of fairness leads to the increased trust in algorithmic decision making. These findings illustrate that trust judgments can be influenced by fairness information which are comprehensively discussed both theoretical and practically in a dedicated section.
% Contributions of this paper include:
% \begin{itemize}
%     \item Propose
%     \item Find
%     \item Get
% \end{itemize}

% Mathematical fairness may be different from human's perception of fairness.
% Invite two groups of participants, one group is on ML technologies, and another group is on only management and no technical backgrounds related to the position. 
% We may then set up a style transfer model to transfer technical group's response style to management group. This allows us to learn how a manager with technical background will response to the recruiting problems.

\section{Related Work}

\subsection{Fairness}
% dwork2018decoupled,dwork2012fairness,
With the increasing uses of AI in critical domains especially human related decision making such as allocation of social benefits, hiring, and criminal justice \cite{berk2018fairness,feldman2015certifying}, fairness is becoming one of key concerns in algorithmic decision making. 
The current research on fairness in machine learning focuses on the formalisation of the definition of fairness and quantifying the unfairness (bias) of an algorithm with different metrics \cite{corbett2018measure,nabi2018fair,glymour2019measuring}. These work typically begins by outlining fairness in the context of different protected attributes (sex, race, origin, culture, etc.) receiving equal treatments by algorithms \cite{kilbertus2017avoiding,bellamy_ai_2018}.
Various definitions on fairness are investigated ranging from statistical bias, group fairness, individual fairness, to process fairness, and others resulting in 21 different definitions \cite{narayanan_translation_2018}. However, it is impossible to satisfy all definitions of fairness at the same time \cite{corbett2017algorithmic,karthik2020impossibility}. Despite the proliferation of fairness definitions and unfairness quantification approaches, little work is found to investigate human's perceived fairness (perception of fairness) when the fairness defined by a specific definition is introduced.

This paper uses statistical parity as the definition of fairness to investigate human perception of fairness in algorithmic decision making.
The statistical parity and its utilisation in this work are briefly described in section~\ref{sec:preliminary}.

% We chose this definition as the mathematically definition of statistical parity allows us to investigate perception of fairness. In other words understand how one's trust is affected by voluntary fairness. 

\subsection{Trust}

Various researches have been investigated to learn user trust variations in algorithmic decision making. Zhou et al. \cite{zhou_be_2015,zhou_measurable_2015} argued that communicating user trust benefits the evaluation of effectiveness of machine learning approaches. Kizilcec \cite{kizilcec_how_2016} proposed that the transparency of algorithm interfaces can promote awareness and foster user trust. It was found that appropriate transparency of algorithms through explanation benefited the user trust. Ribeiro et al. \cite{ribeiro_why_2016} explained predictions by learning an interpretable model locally around the prediction and visualizing importance of the most relevant features to improve user trust in classifications.
Other studies that empirically tested the importance of explanation to users, in various fields, consistently showed that explanations significantly increase users' confidence and trust \cite{Bilgic_explaining_2005,Symeonidis_moviexplain_2009}. 
Zhou et al. \cite{zhou_physiological_2019} investigated the effects of presentation of influence of training data points on predictions to boost user trust and found that 
the presentation of influences of training data points significantly increased the user trust in predictions, but only for training data points with higher influence values under the high model performance condition.
Zhang et al. \cite{zhang_effect_2020} investigated the effect of confidence score and local explanation on trust. It was found that confidence score can help calibrate people's trust in an AI model, but local explanations were not able to create a perceivable effect for trust calibration as expected, maybe because of the experiment design.
In addition, researchers found that user trust had significant correlations with users' experience of system performance \cite{Zhou_human_2018}. Yin et al. \cite{Yin_does_2018} also found that the stated model accuracy had a significant effect on the extent to which people trust the model, suggesting the importance of communication of ML model performance for user trust.

These previous work primarily focuses on the investigation of effects of explanation and model performance on user trust in algorithmic decision making. However, less attention has been paid to the perception of fairness and its effects on trust, which is investigated by this paper.

\subsection{Fairness and Trust}

It was found that perceptions of fair treatment on customers are important in driving trustworthiness and engendering trust in the banking context \cite{roy_impact_2015}.
Earle and Siegrist \cite{earle_relation_2008} found that 
procedural fairness had no effects on trust when issue importance was high, while procedural fairness had moderate effects on perceived fairness and trust when issue importance was low.
Nikbin et al. \cite{nikbin_effects_2011} found that perceived service fairness had a significant relationship with trust, and confirmed the mediating role of satisfaction and trust in the relationship between perceived service fairness and behavioural intention. 
Lee \cite{lee_understanding_2018} investigated how people perceived decisions made by algorithms as compared with decisions made by humans in a management context. It was found that algorithmic decisions were perceived as less fair and trustworthy and evoked more negative emotion than human decisions. 

Previous work pays more attention to relations between the perception of fairness especially procedural fairness and user trust in social interaction context such as marketing and services, however, little work is found on the effects of fairness on user trust in algorithmic decision making. 
This study investigates whether the introduced fairness is positively received by humans and how such fairness affects user trust by simulating a candidate shortlisting in a human resource recruitment setting in algorithmic decision making.

% Algorithmic Mediation in Group Decisions: Fairness Perceptions of Algorithmically Mediated vs. Discussion-Based Social Division \cite{lee_algorithmic_2017}

% The Role of Procedural Fairness in Trust and Trustful Behavior\cite{muller_role_2008}
% \url{https://ideas.repec.org/p/xrs/sfbmaa/08-21.html}

\section{Preliminary Knowledge}
\label{sec:preliminary}

Fairness is a complex and multi-faceted concept that depends on context and culture \cite{bellamy_ai_2018}. Narayanan \cite{narayanan_translation_2018} described at least 21 mathematical definitions of fairness from the literature. This is because of different reasons such as different contexts/applications, different stakeholders, impossibility theorems, as well as allocative versus representational harms. 

% It is also shown that it is impossible to satisfy all definitions of fairness at the same time \cite{bellamy_ai_2018}. 
% From ethical perspective, fairness can refer to \cite{rovatsos_ai_2019}:
% \begin{itemize}
%     \item Equity: use of discretion and fairness when applying justice;
%     \item Social justice: equality and solidarity in a society;
%     \item Distributive justice: appropriateness of the distribution of benefits;
%     \item Procedural justice: appropriateness of procedures used to allocate benefits;
%     \item Interactional justice: appropriateness of interpersonal treatment.
% \end{itemize}

In this study, the statistical parity, one of group fairness definitions, is used to represent fairness. The statistical parity suggests that a predictor is fair if the prediction $\hat{Y}$ is independent of the protected attribute $Z$ so that

\begin{equation}
\label{Statsparity}
    P\left ( \hat{Y}|Z \right ) = P\left ( \hat{Y} \right ).
\end{equation}

It also means that subjects in both protected and unprotected groups have equal probability of being assigned to the positive predicted class. Taken the recruitment as an example, this would imply equal probability for male and female applicants to have positive predicted recruitment: 

\begin{equation}
    P\left ( \hat{Y}=1|Z=0 \right ) = P\left ( \hat{Y}=1|Z=1 \right ) 
\end{equation}

\noindent where $Z=0$ represents male applicants and $Z=1$ represents female applicants. Based on these preliminaries, statistical parity difference ($PD$) is defined as:
\begin{equation}
    PD = \left | P\left ( \hat{Y}=1|Z=0 \right ) -P\left ( \hat{Y}=1|Z=1 \right ) \right |
\end{equation}

\noindent where $PD$ is in the range of $[0,1]$. $PD=0$ represents the complete fairness, and $PD=1$ represents the complete unfairness. This paper manipulates various fairness levels of $PD$ between $[0,1]$ to learn how introduced fairness is perceived and affects trust in algorithmic decision making.

\section{Hypotheses}

% Parity difference is linear, we want to see whether perception of fairness is also linear.

% We also want to find at what parity difference level that perception of fairness will be changed significantly.

This paper poses the following hypotheses:
\begin{itemize}
    \item [H1] The human perceived fairness will be positively related to the introduced fairness. That is, the high level of introduced fairness will result in the high level of perceived fairness by humans, and vice versa; 
    \item [H2] The user trust will be positively related to the introduced fairness. That is, the high level of introduced fairness will result in the high level of trust in algorithmic decision making, and vice versa; 
    \item [H3] Humans will be more sensitive to the changes of high level of introduced fairness than the low level of introduced fairness. 
\end{itemize}

\section{Method}

\subsection{Case Study}
% We are interested in measuring how fair you believe AI algorithms are and what the trust level is when used to make decisions in our daily lives, such as recruiting candidates for positions from a large number of applications. 

In this study, a company is supposed to recruit staff for a position. The company posted the job description and different applicants submitted their applications. Participants are supposed to be a recruiter to shortlist applicants for the position. A machine learning system named Automatic Recruiting Assistant (ARA) is used to help participants make decisions on the shortlisting of applicants. ARA is a candidate assessment software that uses historical recruiting data to train a machine learning model and predict whether a candidate will be shortlisted. The ARA’s accuracy keeps constant for all shortlisting.

% Put all information in a formatted profile. See the reference [3].

\subsection{Fairness Scale}
% Overall fairness does not consider the distinctions among procedural, distributive, interpersonal, and informational facets.

This study investigates the perception of fairness, which is measured with a single scale that focuses on a global perception of appropriateness \cite{colquitt_measuring_2015}. In this study, the questionnaire on the fairness is ``overall, female and male applicants are treated fairly by ARA''. The scale is on a 5-point Likert-type response scale ranging from 1 (strongly disagree) to 5 (strongly agree).

\subsection{Trust Scale}
% merritt_affective_2011,
Trust is assessed with six items using self-report scales as the following  
\cite{merritt_i_2013}. The scale is on a 5-point Likert-type response scale ranging from 1 (strongly disagree) to 5 (strongly agree).

\begin{itemize}
\item I believe the ARA is a competent performer.
\item I trust the ARA.
\item I have confidence in the advice given by the ARA.
\item I can depend on the ARA.
\item I can rely on the ARA to behave in consistent ways.
\item I can rely on the ARA to do its best every time I take its advice.
\end{itemize}

\section{Experiment}
This section details the experiment to examine our hypotheses on the introduced fairness, human perception of fairness and trust in an automatic recruitment algorithm.

\subsection{Task Design}
Tasks were designed to investigate effects of different fairness levels on user trust in algorithmic decision making. 
The protected attribute in this study is the gender of applicants.
In this case, the $PD$ is the difference of shortlisted rate by the gender.
In this study, fairness was introduced by manipulating $PD$ with its discrete values of 0, 0.1, 0.2, 0.3, 0.4, …, 0.8, 0.9, and 1.0, where each $PD$'s discrete value was used as a measure of fairness to define the number of male and female applicants as well as number of male and female applicants shortlisted in each task respectively. Table~\ref{tab:tasks} shows 11 task examples corresponding to different $PD$ values. In this table, ``Rate (Male)" represents the predicted success rate for male applicants, ``Rate (Female)" represents the predicted success rate for female applicants, ``Male \#'' represents the number of male applicants, ``Female \#'' represents the number of female applicants, ``Listed Male \#'' represents the number of shortlisted male applicants, and ``Listed Female \#'' represents the number of shortlisted female applicants. With the same settings of $PD$ as in the table, different number of male and female applicants were used to generate another 11 tasks. 
All together 22 tasks were conducted by each participant. Two additional training tasks were also conducted by each participant before the formal tasks. The order of tasks was randomized during the experiment to avoid any bias.

\begin{table*}[!htb]
  \caption{Experiment tasks.}
    \centering
    \begin{tabular}{l|l|l|l|l|l|l|l}
    \hline
    Task$\#$ & $PD$ & Rate (Male) & Rate (Female) & Male$\#$ & Female$\#$ & Listed Male$\#$ &Listed Female$\#$  \\
    \hline
        1  &  0 & 0.8 & 0.8 & 10 & 10 & 8 & 8\\
         2	& 0.1&	0.7&	0.8&	10&	5&	7&	4\\
        3&	0.2&	0.6&	0.8&	5&	5&	3&	4\\
        4&	0.3&	0.8&	0.5&	5&	10&	4&	5\\
        5&	0.4&	0.8&	0.4&	5&	5&	4&	2\\
        6&	0.5&	0.7&	0.2&	10&	5&	7&	1\\
        7&	0.6&	0.8&	0.2&	5&	5&	4&	1\\
        8&	0.7&	0.1&	0.8&	10&	5&	1&	4\\
        9&	0.8&	0.9&	0.1&	10&	10&	9&	1\\
        10&	0.9&	0.1&	1&	10&	10&	1&	10\\
        11&	1&	1&	0&	5&	10&	5&	0\\
         \hline
    \end{tabular}
    \label{tab:tasks}
\end{table*}

% In this experiment, for each task we have different number of applicants who applied and the corresponding number of applicants who were finally shortlisted. Further we introduced gender diversity by displaying the numbers separately for males and females. By introducing such disparity we try to influence the user perception of fairness and trust in each task. In total we have designed 22 tasks for each applicant to undertake in one sitting and the time from start to submit is also recorded as a data point to be considered while analysing the final data collected.

\begin{figure}[!htb]
  \centering
  \includegraphics[width=0.99\linewidth]{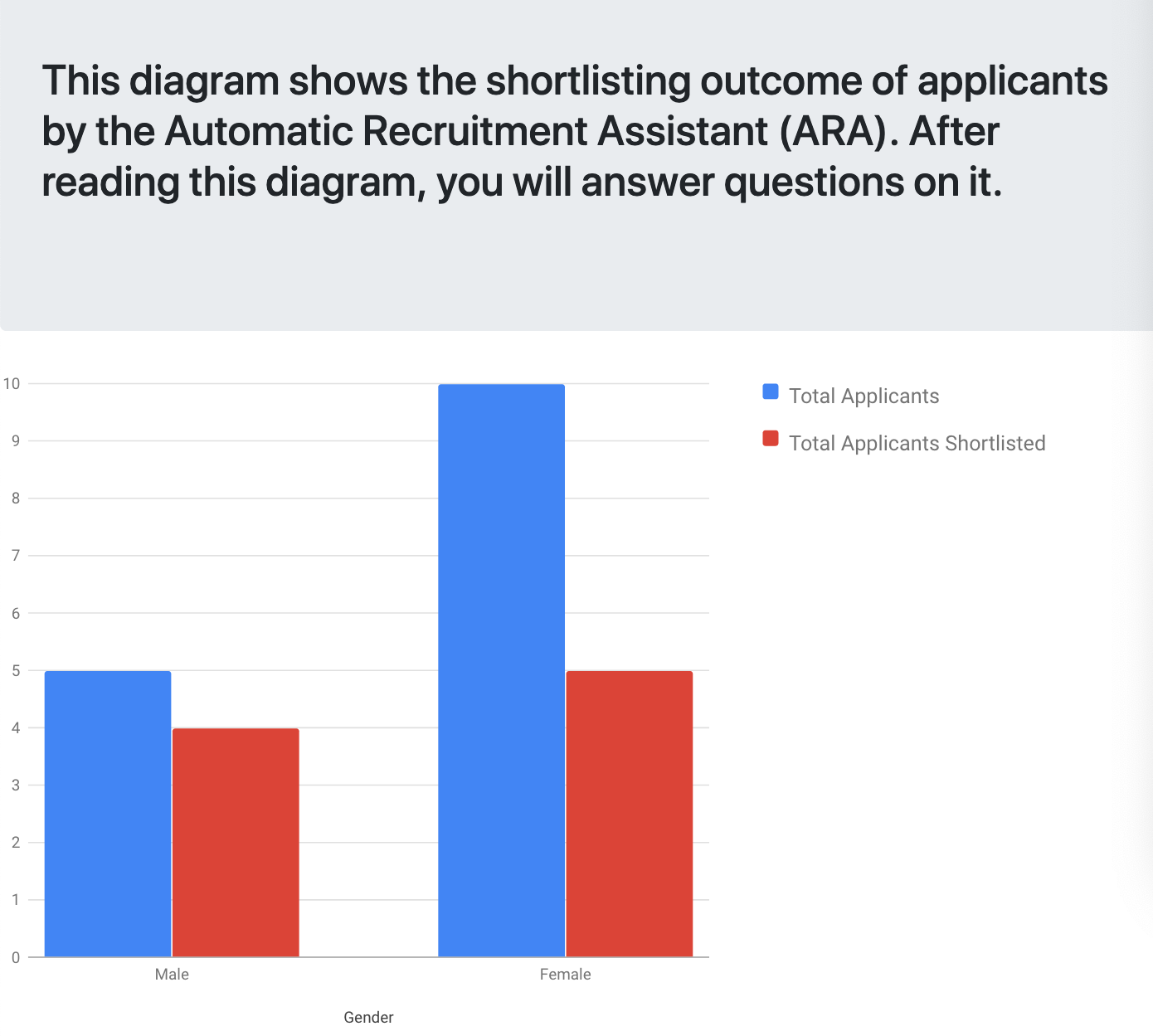}
    \caption{Screenshot of the experiment}
\label{fig:screenshot}       
\end{figure}

\subsection{Experiment Setup}

Due to social distancing restrictions and lockdown policies during the COVID-19 pandemic, our experiment was implemented using the flask framework in Python and was deployed on the Heroku cloud server online. The deployed application link was then shared with participants to invite them to conduct tasks. In this study, participant responses to tasks were stored in a MySQL database that was directly connected to the flask application. Figure~\ref{fig:screenshot} shows the screenshot of a task conducted in the experiment. 

% We used gender diversity as a key differentiator to gauge the perception of fairness and trust. 
% the questionnaire application was prepared 

\subsection{Participants and Data Collection}

20 participants were invited via various means of communications such as emails, text messages and social media posts who are mainly university students around the age group of 20-30 years with the average of around 25 years old. After each task was displayed on the screen, the participants were asked to answer seven questions based on the task. The first question was on fairness of applicant shortlisting shown in the task while the other six questions were on the trust of the participant in the decision making from the ARA.

\section{Results}

This study aims to understand: 1) how the introduced fairness is perceived by humans, and 2) how the introduced fairness affects user trust in algorithmic decision making. In order to perform the analyses, we first normalised the collected trust and fairness data. We then performed one-way ANOVA tests on the normalised data followed by post-hoc comparison using Tukey HSD tests. 
The fairness and trust values were normalised with respect to each subject to minimise individual differences in rating behavior using the equation given below:

\begin{equation}
	T_i^N = \frac{T_i - T_i^{min}}{T_i^{max} - T_i^{min}}
\end{equation}
\noindent where $T_i$ and $T_i^N$  are the original fairness or trust ratings and the normalised fairness or trust rating respectively from the user $i$, $T_i^{min}$ and $T_i^{max}$ are the minimum and maximum of the ratings respectively from the user $i$ in all of his/her tasks.

\subsection{Perception of Fairness}

\begin{figure}[!htb]
  \centering
  \includegraphics[width=0.99\linewidth]{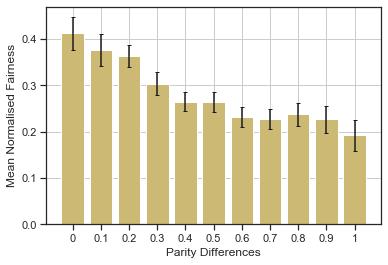}
    \caption{Mean normalised perceived fairness over introduced fairness.}
\label{fig:fairness_result}       
\end{figure}

Figure~\ref{fig:fairness_result} shows the mean normalised perceived fairness (perception of fairness) over introduced fairness (error bars represent the 95\% confidence interval of a mean and it is the same in other figures). A one-way ANOVA test found that there were statistically significant differences in perceived fairness among 11 introduced fairness levels ($F(10, 429) = 29.872, p<.000$).
The further post-hoc comparison with Tukey HSD tests were conducted to test pair-wised differences in perceived fairness between two introduced fairness levels. It was found that the perceived fairness at $PD=0, 0.1,$ and $0.2$ had significant differences with all other $PD$ levels from 0.4 to 1.0 respectively (for all, $p<.001$). The perceived fairness at $PD=0$ ($p<.001$) and $0.1$ ($p<.005$) also had significant differences with $PD=0.3$ respectively. However, there were no significant differences found in perceived fairness among any pair of $PD$ at $0, 0.1,$ and $0.2$. It was also found that the perceived fairness at $PD=0.3$ had significant differences with $PD=0.6, 0.7,...,1.0$ respectively (for all, $p<.017$). Furthermore, the perceived fairness at $PD=0.4$ ($p<.006$) and $0.5$ ($p<.005$) had significant differences with $PD=1.0$ respectively. 
Despite no other significant difference found in perceived fairness among introduced fairness levels, 
Figure~\ref{fig:fairness_result} shows that the perceived fairness has a clear decreasing trend with the decrease of introduced fairness (increase of $PD$ levels). The results suggest that participants' perception of fairness was positively related to the introduced fairness (H1), but was not sensitive to the small changes of introduced fairness. 
Moreover, participants were more sensitive to the perceived fairness with high levels than low levels as we expected (H3).
These findings also imply that the introduced fairness can be safely used to validate the perception of fairness of humans.

Following the findings of the trend of perceived fairness as described above, we divided introduced fairness into three groups: 

\begin{itemize}
    \item Group A (high level of introduced fairness group): $PD=0, 0.1, 0.2, 0.3$;
    \item Group B (middle level of introduced fairness group): $PD$=0.4, 0.5, 0.6, 0.7;
    \item Group C (low level of introduced fairness group): $PD=0.8, 0.9, 1.0$.
\end{itemize}

\begin{figure}[!htb]
  \centering
  \includegraphics[width=0.9\linewidth]{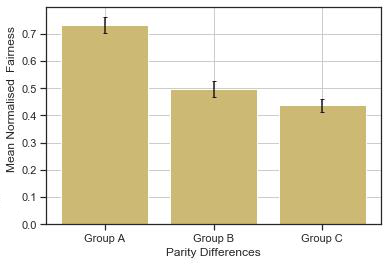}
    \caption{Mean normalised perceived fairness over grouped introduced fairness.}
\label{fig:group_fairness_result}       
\end{figure}

Therefore, a one-way ANOVA test was conducted for these three groups of introduced fairness. It was found that there were statistically significant differences in perceived fairness among three introduced fairness levels ($F(2, 117) = 104.725, p<.000$).
The post-hoc comparison with Tukey HSD tests were conducted to test pair-wised differences in perceived fairness between two introduced fairness group levels. It was found that the perceived fairness at the introduced fairness level of Group A was significantly higher than that at levels of Group B ($p<.001$) and Group C ($p<.001$) respectively. The perceived fairness at the introduced fairness level of Group B was also significantly higher than that at the level of Group C ($p<.001$).
The results show that human perception of fairness was positively related to the introduced fairness. The findings imply that the introduced fairness based on $PD$ can safely reflect perception of fairness in algorithmic decision making.

% Parity difference is linear, we want to see whether perception of fairness is also linear.

% Perception of fairness is not linear.

\subsection{Fairness and Trust}

\begin{figure}[!htb]
  \centering
  \includegraphics[width=0.99\linewidth]{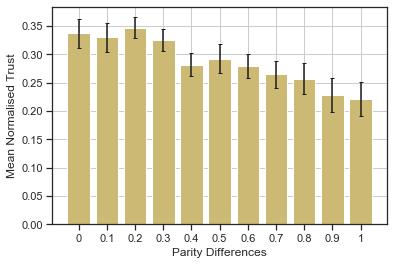}
    \caption{Mean normalised trust over fairness.}
\label{fig:trust_result}       
\end{figure}

Figure~\ref{fig:trust_result} shows mean normalised trust ratings over introduced fairness ($PD$) levels. 
A one-way ANOVA test found that there were statistically significant differences in trust ratings among 11 fairness levels ($F(10, 429) = 11.550, p<.000$). Then the post-hoc comparison using Tukey HSD tests found significant differences in trust responses between introduced fairness level pairs as shown in Table~\ref{tab:trust_posthoc}. It shows that participants had significantly higher trust in AI-informed decisions under high introduced fairness levels (low $PD$ values) than that under low introduced fairness levels (high $PD$ values). For example, participants had significantly higher trust under $PD=0$ than that under $PD=0.7$, $p<.003$. However, user trust did not show significant differences under high introduced fairness levels (e.g. $PD$ = 0, 0.1, 0.2, 0.3).

\begin{table}[!htb]
  \caption{Post-hoc comparison using Tukey HSD tests on trust between introduced fairness level pairs.}
    \centering
    \begin{tabular}{l|l|l V{3} l|l|l}
    \hline
    PD1 & PD2 & $p$-value & PD1 & PD2 & $p$-value  \\
    \hline
    0  & 0.7 & .003 &   0.1  & 0.7 & .015\\
    0  & 0.8 & .001 &   0.1  & 0.8 & .003\\
    0  & 0.9 & .001 &   0.1  & 0.9 & .001\\
    0  & 1.0 & .001 &   0.1  & 1.0 & .001\\
    \hline
    0.2  & 0.4 & .014 &   0.3  & 0.7 & .032\\
    0.2  & 0.6 & .010 &   0.3  & 0.8 & .008\\
    0.2  & 0.7 & .001 &   0.3  & 0.9 & .001\\
    0.2  & 0.8 & .001 &   0.3  & 1.0 & .001\\
    \cline{4-6}
    0.2  & 0.9 & .001 &    0.5  & 0.9 & .018\\
    0.2  & 1.0 & .001 &   0.5  & 1.0 & .004\\
    \hline
    0.4  & 1.0 & .032 & 0.6  & 1.0 & .046\\
    \hline
    \end{tabular}
    \label{tab:trust_posthoc}
\end{table}

\begin{figure}[!htb]
  \centering
  \includegraphics[width=0.99\linewidth]{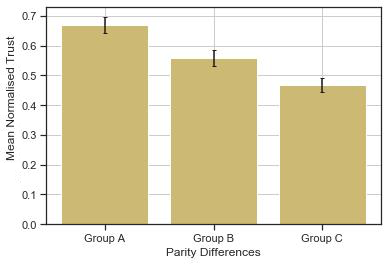}
    \caption{Mean normalised trust over grouped introduced fairness.}
\label{fig:group_trust_result}       
\end{figure}

We further analyse trust differences under three introduced fairness group levels of A, B, C. as described above, 
a one-way ANOVA test found that there were statistically significant differences in user trust among three introduced fairness group levels ($F(2, 117) = 48.272, p<.000$).
The further post-hoc comparison with Tukey HSD tests found that user trust was significantly higher at the introduced fairness level of Group A than that at the levels of Group B ($p<.001$) and Group C ($p<.001$) respectively. User trust was also significantly higher at the introduced fairness level of Group B than that at the level of Group C ($p<.001$). 

The findings suggest that user trust had a positive relationship with the introduced fairness as we expected (H2). The higher the introduced fairness level was, the higher trust in decisions users had.

\section{Discussion}

Our study found that the introduced fairness was positively related to the perceived fairness by humans. Besides, it also shows that high levels of introduced fairness resulted in high levels of human perception of fairness. These findings confirm that the introduced fairness level can be safely used to evaluate the human perception of fairness. Furthermore, the introduced fairness was also positively related to user's trust in algorithmic decision making. Once again we see that the high level of introduced fairness benefited user trust.  
It was also found that participants were more sensitive to the introduced fairness with high levels than low levels.

Fairness heuristic theory \cite{lind_fairness_2001,van_den_bos_uncertainty_2001} suggests that when individuals face uncertain circumstances they rely on impressions of fairness to determine whether to cooperate and enter into exchange relationships with the other party, which suggests that individuals use fairness judgements to form their perceptions of trust.
The social exchange theory \cite{blau_exchange_1964} also argues that fair actions and the treatment by one party generate reciprocation in the form of trust by the other party in the exchange.
In the context of candidate shortlisting in human resource settings utilised in this paper, recruiters were unsure about the outcomes from the Automatic Recruiting Assistant. As a result, recruiters formed the trust perception based upon fairness perception, and high level of perception of fairness resulted in the high level of trust in algorithmic decision making. 

These findings have significant implications in algorithmic decision making applications. For example, when the trust is difficult to examine in algorithmic decision making, human perception of fairness can be used to estimate user trust in algorithmic decision making. While human perception of fairness is positively related to introduced fairness. Our findings also imply that the introduced fairness can be safely used to validate the human perception of fairness.

Furthermore, since human is more sensitive to the high level of fairness, the high level of fairness instead of the low level of fairness can be explicitly presented in the user interface of AI applications to boost user trust in algorithmic decision making.

Overall, the findings from this study at least have the following implications: 1) the estimation of user trust in algorithmic decision making by human perception of fairness; 2) the user interface design of AI applications to boost user trust by explicitly presenting high level of fairness to users; 3) manipulation of human perception of fairness by manipulating level of introduced fairness.

% This study investigates whether fairness causally determines trust. 
% To directly address the causal link between fairness and trust, we manipulate perceived fairness by way of a new paradigm.

% \url{https://arxiv.org/pdf/1701.08230.pdf}
% Algorithms have the potential to improve the efficiency and equity of decisions, but their design and application raise complex questions for researchers and policymakers. By clarifying the implications of competing notions of algorithmic fairness, we hope our analysis fosters discussion and informs policy.

% Trust is identified as an outcome of organisational fairness in sociology and organisational behaviour research \cite{lind_social_1988,aryee_trust_2002}. 

%\cite{ferreira-oliveira_decision_2018}

% on fair theory and trust
%https://www-tandfonline-com.ezproxy.lib.uts.edu.au/doi/pdf/10.1080/0267257X.2015.1036101?needAccess=true&

\section{Conclusion and Future Work}

Fairness is a key concern in algorithmic decision making in many application areas including shortlisting candidates for advertised positions or loan approvals based on historical records. The human perception of fairness affects adoptions of AI applications.
This paper understood the relations between the introduced fairness and human perception of fairness and investigated how the introduced fairness affected user trust in algorithmic decision making. Experimental results showed that the introduced fairness was positively related to human perception of fairness, and concurrently it was also positively related to user's trust. Interestingly, the users were more sensitive to fairness with high levels than that with low levels. The findings can be used to help to estimate trust in algorithmic decision making and user interface design for AI solutions. The future work of this study will focus on the introduction of AI explanations into the pipeline to understand their effects on user trust in algorithmic decision making.

% The area of transparency, or explainability, has emerged as a way to aid our understanding of the inner workings of a machine learning model.
% Fairness is a key concern in many application areas including selecting candidates for hire or approving loans in banking. A popular family of approaches for transparency provide feature importance, or saliency, scores for a given input. These scores show how important each feature of the input was to the algorithm’s decision locally around the input.
% \cite{dimanov_you_2020}

\begin{acks}
Authors would like to thank all participants for this study.
\end{acks}

\bibliographystyle{ACM-Reference-Format}
\bibliography{fairness,tml}


\begin{thebibliography}{41}

%%% the Web address.

\ifx \showCODEN    \undefined \def \showCODEN     #1{\unskip}     \fi
\ifx \showDOI      \undefined \def \showDOI       #1{#1}\fi
\ifx \showISBNx    \undefined \def \showISBNx     #1{\unskip}     \fi
\ifx \showISBNxiii \undefined \def \showISBNxiii  #1{\unskip}     \fi
\ifx \showISSN     \undefined \def \showISSN      #1{\unskip}     \fi
\ifx \showLCCN     \undefined \def \showLCCN      #1{\unskip}     \fi
\ifx \shownote     \undefined \def \shownote      #1{#1}          \fi
\ifx \showarticletitle \undefined \def \showarticletitle #1{#1}   \fi
\ifx \showURL      \undefined \def \showURL       {\relax}        \fi
% invisible to TeX
\providecommand\bibfield[2]{#2}
\providecommand\bibinfo[2]{#2}
\providecommand\natexlab[1]{#1}
\providecommand\showeprint[2][]{arXiv:#2}

\bibitem[\protect\citeauthoryear{Aggarwal and Larrick}{Aggarwal and
  Larrick}{2012}]%
        {aggarwal_when_2012}
\bibfield{author}{\bibinfo{person}{Pankaj Aggarwal} {and}
  \bibinfo{person}{Richard~P. Larrick}.} \bibinfo{year}{{2012}}\natexlab{}.
\newblock \showarticletitle{{When consumers care about being treated fairly:
  The interaction of relationship norms and fairness norms}}.
\newblock \bibinfo{journal}{\emph{{Journal of Consumer Psychology}}}
  \bibinfo{volume}{{22}}, \bibinfo{number}{{1, SI}} (\bibinfo{year}{{2012}}),
  \bibinfo{pages}{{114--127}}.
\newblock


\bibitem[\protect\citeauthoryear{Bellamy and {et al.}}{Bellamy and {et
  al.}}{2018}]%
        {bellamy_ai_2018}
\bibfield{author}{\bibinfo{person}{Rachel K.~E. Bellamy} {and}
  \bibinfo{person}{{et al.}}} \bibinfo{year}{2018}\natexlab{}.
\newblock \showarticletitle{{AI} Fairness 360: An Extensible Toolkit for
  Detecting, Understanding, and Mitigating Unwanted Algorithmic Bias}.
\newblock \bibinfo{journal}{\emph{{arXiv}:1810.01943 [cs]}}
  (\bibinfo{year}{2018}).
\newblock
\showeprint[arxiv]{1810.01943}


\bibitem[\protect\citeauthoryear{Berk, Heidari, Jabbari, Kearns, and Roth}{Berk
  et~al\mbox{.}}{2018}]%
        {berk2018fairness}
\bibfield{author}{\bibinfo{person}{Richard Berk}, \bibinfo{person}{Hoda
  Heidari}, \bibinfo{person}{Shahin Jabbari}, \bibinfo{person}{Michael Kearns},
  {and} \bibinfo{person}{Aaron Roth}.} \bibinfo{year}{2018}\natexlab{}.
\newblock \showarticletitle{Fairness in criminal justice risk assessments: The
  state of the art}.
\newblock \bibinfo{journal}{\emph{Sociological Methods \& Research}}
  (\bibinfo{year}{2018}), \bibinfo{pages}{0049124118782533}.
\newblock


\bibitem[\protect\citeauthoryear{Bilgic and Mooney}{Bilgic and Mooney}{2005}]%
        {Bilgic_explaining_2005}
\bibfield{author}{\bibinfo{person}{Mustafa Bilgic} {and}
  \bibinfo{person}{Raymond Mooney}.} \bibinfo{year}{2005}\natexlab{}.
\newblock \showarticletitle{Explaining Recommendations: Satisfaction vs.
  Promotion}. In \bibinfo{booktitle}{\emph{Proceedings of Beyond
  Personalization 2005: A Workshop on the Next Stage of Recommender Systems
  Research at 2005 IUI}}.
\newblock


\bibitem[\protect\citeauthoryear{Blau}{Blau}{1964}]%
        {blau_exchange_1964}
\bibfield{author}{\bibinfo{person}{Peter~M. Blau}.}
  \bibinfo{year}{1964}\natexlab{}.
\newblock \bibinfo{booktitle}{\emph{Exchange and Power in Social Life}}.
\newblock \bibinfo{publisher}{John Wiley and Sons}, \bibinfo{address}{New York,
  NY}.
\newblock


\bibitem[\protect\citeauthoryear{Chen and Zhou}{Chen and Zhou}{2019}]%
        {zhou_ai_2019}
\bibfield{author}{\bibinfo{person}{Fang Chen} {and} \bibinfo{person}{Jianlong
  Zhou}.} \bibinfo{year}{2019}\natexlab{}.
\newblock \showarticletitle{{AI} in the public interest}.
\newblock In \bibinfo{booktitle}{\emph{Closer to the Machine: Technical,
  Social, and Legal Aspects of {AI}}}, \bibfield{editor}{\bibinfo{person}{Cliff
  Bertram}, \bibinfo{person}{Asher Gibson}, {and} \bibinfo{person}{Adriana
  Nugent}} (Eds.). \bibinfo{publisher}{Office of the Victorian Information
  Commissioner}.
\newblock


\bibitem[\protect\citeauthoryear{Colquitt and Rodell}{Colquitt and
  Rodell}{2015}]%
        {colquitt_measuring_2015}
\bibfield{author}{\bibinfo{person}{Jason~A. Colquitt} {and}
  \bibinfo{person}{Jessica~B. Rodell}.} \bibinfo{year}{2015}\natexlab{}.
\newblock \showarticletitle{Measuring Justice and Fairness}.
\newblock In \bibinfo{booktitle}{\emph{The Oxford Handbook of Justice in the
  Workplace}}, \bibfield{editor}{\bibinfo{person}{Russell~S. Cropanzano} {and}
  \bibinfo{person}{Maureen~L. Ambrose}} (Eds.). \bibinfo{publisher}{Oxford
  University Press}.
\newblock
\showISBNx{978-0-19-998141-0}


\bibitem[\protect\citeauthoryear{Corbett-Davies and Goel}{Corbett-Davies and
  Goel}{2018}]%
        {corbett2018measure}
\bibfield{author}{\bibinfo{person}{Sam Corbett-Davies} {and}
  \bibinfo{person}{Sharad Goel}.} \bibinfo{year}{2018}\natexlab{}.
\newblock \showarticletitle{The measure and mismeasure of fairness: A critical
  review of fair machine learning}.
\newblock \bibinfo{journal}{\emph{arXiv preprint arXiv:1808.00023}}
  (\bibinfo{year}{2018}).
\newblock


\bibitem[\protect\citeauthoryear{Corbett-Davies, Pierson, Feller, Goel, and
  Huq}{Corbett-Davies et~al\mbox{.}}{2017}]%
        {corbett2017algorithmic}
\bibfield{author}{\bibinfo{person}{Sam Corbett-Davies}, \bibinfo{person}{Emma
  Pierson}, \bibinfo{person}{Avi Feller}, \bibinfo{person}{Sharad Goel}, {and}
  \bibinfo{person}{Aziz Huq}.} \bibinfo{year}{2017}\natexlab{}.
\newblock \showarticletitle{Algorithmic decision making and the cost of
  fairness}. In \bibinfo{booktitle}{\emph{Proceedings of the 23rd acm sigkdd
  international conference on knowledge discovery and data mining}}.
  \bibinfo{pages}{797--806}.
\newblock


\bibitem[\protect\citeauthoryear{Earle and Siegrist}{Earle and
  Siegrist}{2008}]%
        {earle_relation_2008}
\bibfield{author}{\bibinfo{person}{Timothy~C. Earle} {and}
  \bibinfo{person}{Michael Siegrist}.} \bibinfo{year}{2008}\natexlab{}.
\newblock \showarticletitle{On the Relation Between Trust and Fairness in
  Environmental Risk Management}.
\newblock \bibinfo{journal}{\emph{Risk Analysis}} \bibinfo{volume}{28},
  \bibinfo{number}{5} (\bibinfo{date}{October} \bibinfo{year}{2008}),
  \bibinfo{pages}{1395--1414}.
\newblock


\bibitem[\protect\citeauthoryear{Feldman, Friedler, Moeller, Scheidegger, and
  Venkatasubramanian}{Feldman et~al\mbox{.}}{2015}]%
        {feldman2015certifying}
\bibfield{author}{\bibinfo{person}{Michael Feldman}, \bibinfo{person}{Sorelle~A
  Friedler}, \bibinfo{person}{John Moeller}, \bibinfo{person}{Carlos
  Scheidegger}, {and} \bibinfo{person}{Suresh Venkatasubramanian}.}
  \bibinfo{year}{2015}\natexlab{}.
\newblock \showarticletitle{Certifying and removing disparate impact}. In
  \bibinfo{booktitle}{\emph{Proceedings of KDD2015}}.
  \bibinfo{pages}{259--268}.
\newblock


\bibitem[\protect\citeauthoryear{Glymour and Herington}{Glymour and
  Herington}{2019}]%
        {glymour2019measuring}
\bibfield{author}{\bibinfo{person}{Bruce Glymour} {and}
  \bibinfo{person}{Jonathan Herington}.} \bibinfo{year}{2019}\natexlab{}.
\newblock \showarticletitle{Measuring the biases that matter: The ethical and
  casual foundations for measures of fairness in algorithms}. In
  \bibinfo{booktitle}{\emph{Proceedings of the conference on fairness,
  accountability, and transparency}}. \bibinfo{pages}{269--278}.
\newblock


\bibitem[\protect\citeauthoryear{Gugnani and Misra}{Gugnani and Misra}{2020}]%
        {gugnani_implicit_2020}
\bibfield{author}{\bibinfo{person}{Akshay Gugnani} {and}
  \bibinfo{person}{Hemant Misra}.} \bibinfo{year}{2020}\natexlab{}.
\newblock \showarticletitle{Implicit Skills Extraction Using Document Embedding
  and Its Use in Job Recommendation}. In \bibinfo{booktitle}{\emph{Proceedings
  of the {AAAI} Conference on Artificial Intelligence}},
  Vol.~\bibinfo{volume}{34}. \bibinfo{pages}{13286--13293}.
\newblock


\bibitem[\protect\citeauthoryear{Hughes, Robert, Frady, and Arroyos}{Hughes
  et~al\mbox{.}}{2019}]%
        {hughes_artificial_2019}
\bibfield{author}{\bibinfo{person}{Claretha Hughes}, \bibinfo{person}{Lionel
  Robert}, \bibinfo{person}{Kris Frady}, {and} \bibinfo{person}{Adam Arroyos}.}
  \bibinfo{year}{2019}\natexlab{}.
\newblock \showarticletitle{Artificial Intelligence, Employee Engagement,
  Fairness, and Job Outcomes}.
\newblock In \bibinfo{booktitle}{\emph{Managing Technology and Middle- and
  Low-skilled Employees}}. \bibinfo{pages}{61--68}.
\newblock


\bibitem[\protect\citeauthoryear{Karthik et~al\mbox{.}}{Karthik
  et~al\mbox{.}}{2020}]%
        {karthik2020impossibility}
\bibfield{author}{\bibinfo{person}{S Karthik} {et~al\mbox{.}}}
  \bibinfo{year}{2020}\natexlab{}.
\newblock \showarticletitle{The Impossibility Theorem of Machine Fairness--A
  Causal Perspective}.
\newblock \bibinfo{journal}{\emph{arXiv e-prints}} (\bibinfo{year}{2020}),
  \bibinfo{pages}{arXiv--2007}.
\newblock


\bibitem[\protect\citeauthoryear{Kilbertus, Carulla, Parascandolo, Hardt,
  Janzing, and Sch{\"o}lkopf}{Kilbertus et~al\mbox{.}}{2017}]%
        {kilbertus2017avoiding}
\bibfield{author}{\bibinfo{person}{Niki Kilbertus},
  \bibinfo{person}{Mateo~Rojas Carulla}, \bibinfo{person}{Giambattista
  Parascandolo}, \bibinfo{person}{Moritz Hardt}, \bibinfo{person}{Dominik
  Janzing}, {and} \bibinfo{person}{Bernhard Sch{\"o}lkopf}.}
  \bibinfo{year}{2017}\natexlab{}.
\newblock \showarticletitle{Avoiding discrimination through causal reasoning}.
  In \bibinfo{booktitle}{\emph{Advances in Neural Information Processing
  Systems}}. \bibinfo{pages}{656--666}.
\newblock


\bibitem[\protect\citeauthoryear{Kizilcec}{Kizilcec}{2016}]%
        {kizilcec_how_2016}
\bibfield{author}{\bibinfo{person}{Ren{\'e}~F. Kizilcec}.}
  \bibinfo{year}{2016}\natexlab{}.
\newblock \showarticletitle{How {Much} {Information}?: {Effects} of
  {Transparency} on {Trust} in an {Algorithmic} {Interface}}. In
  \bibinfo{booktitle}{\emph{Proceedings of {CHI}2016}}.
  \bibinfo{pages}{2390--2395}.
\newblock


\bibitem[\protect\citeauthoryear{Komodromos}{Komodromos}{2014}]%
        {komodromos_employees_2014}
\bibfield{author}{\bibinfo{person}{Marcos Komodromos}.}
  \bibinfo{year}{2014}\natexlab{}.
\newblock \showarticletitle{Employees' Perceptions of Trust, Fairness, and the
  Management of Change in Three Private Universities in Cyprus}.
\newblock \bibinfo{journal}{\emph{Journal of Human Resources Management and
  Labor Studies}} \bibinfo{volume}{2}, \bibinfo{number}{2}
  (\bibinfo{date}{July} \bibinfo{year}{2014}), \bibinfo{pages}{35--54}.
\newblock


\bibitem[\protect\citeauthoryear{Lavelle, {McMahan}, and Harris}{Lavelle
  et~al\mbox{.}}{2009}]%
        {lavelle_fairness_2009}
\bibfield{author}{\bibinfo{person}{James~J. Lavelle}, \bibinfo{person}{Gary~C.
  {McMahan}}, {and} \bibinfo{person}{Christopher~M. Harris}.}
  \bibinfo{year}{2009}\natexlab{}.
\newblock \showarticletitle{Fairness in human resource management, social
  exchange relationships, and citizenship behavior: testing linkages of the
  target similarity model among nurses in the United States}.
\newblock \bibinfo{journal}{\emph{The International Journal of Human Resource
  Management}} \bibinfo{volume}{20}, \bibinfo{number}{12} (\bibinfo{date}{12}
  \bibinfo{year}{2009}), \bibinfo{pages}{2419--2434}.
\newblock
\showISSN{0958-5192}


\bibitem[\protect\citeauthoryear{Lee}{Lee}{2018}]%
        {lee_understanding_2018}
\bibfield{author}{\bibinfo{person}{Min~Kyung Lee}.}
  \bibinfo{year}{2018}\natexlab{}.
\newblock \showarticletitle{Understanding perception of algorithmic decisions:
  Fairness, trust, and emotion in response to algorithmic management}.
\newblock \bibinfo{journal}{\emph{Big Data \& Society}} \bibinfo{volume}{5},
  \bibinfo{number}{1} (\bibinfo{date}{June} \bibinfo{year}{2018}),
  \bibinfo{pages}{205395171875668}.
\newblock


\bibitem[\protect\citeauthoryear{Lee, Jain, Cha, Ojha, and Kusbit}{Lee
  et~al\mbox{.}}{2019}]%
        {lee_procedural_2019}
\bibfield{author}{\bibinfo{person}{Min~Kyung Lee}, \bibinfo{person}{Anuraag
  Jain}, \bibinfo{person}{Hea~Jin Cha}, \bibinfo{person}{Shashank Ojha}, {and}
  \bibinfo{person}{Daniel Kusbit}.} \bibinfo{year}{2019}\natexlab{}.
\newblock \showarticletitle{Procedural Justice in Algorithmic Fairness:
  Leveraging Transparency and Outcome Control for Fair Algorithmic Mediation}.
\newblock \bibinfo{journal}{\emph{Proceedings of the {ACM} on Human-Computer
  Interaction}}  \bibinfo{volume}{3} (\bibinfo{date}{November}
  \bibinfo{year}{2019}), \bibinfo{pages}{1--26}.
\newblock
Issue {CSCW}.


\bibitem[\protect\citeauthoryear{Leventhal}{Leventhal}{1980}]%
        {leventhal_what_1980}
\bibfield{author}{\bibinfo{person}{Gerald~S. Leventhal}.}
  \bibinfo{year}{1980}\natexlab{}.
\newblock \showarticletitle{What Should Be Done with Equity Theory?}
\newblock In \bibinfo{booktitle}{\emph{Social Exchange: Advances in Theory and
  Research}}, \bibfield{editor}{\bibinfo{person}{Kenneth~J. Gergen},
  \bibinfo{person}{Martin~S. Greenberg}, {and} \bibinfo{person}{Richard~H.
  Willis}} (Eds.). \bibinfo{publisher}{Springer {US}}, \bibinfo{pages}{27--55}.
\newblock
\showISBNx{978-1-4613-3087-5}


\bibitem[\protect\citeauthoryear{Lind}{Lind}{2001}]%
        {lind_fairness_2001}
\bibfield{author}{\bibinfo{person}{E. Lind}.} \bibinfo{year}{2001}\natexlab{}.
\newblock \showarticletitle{Fairness heuristic theory: Justice judgments as
  pivotal cognitions in organizational relations}. In
  \bibinfo{booktitle}{\emph{Advances in organizational justice}}.
  \bibinfo{publisher}{Stanford University Press}, \bibinfo{pages}{56--88}.
\newblock


\bibitem[\protect\citeauthoryear{Mayer, Davis, and Schoorman}{Mayer
  et~al\mbox{.}}{1995}]%
        {mayer_integrative_1995}
\bibfield{author}{\bibinfo{person}{Roger~C. Mayer}, \bibinfo{person}{James~H.
  Davis}, {and} \bibinfo{person}{F.~David Schoorman}.}
  \bibinfo{year}{1995}\natexlab{}.
\newblock \showarticletitle{An Integrative Model of Organizational Trust}.
\newblock \bibinfo{journal}{\emph{The Academy of Management Review}}
  \bibinfo{volume}{20}, \bibinfo{number}{3} (\bibinfo{date}{July}
  \bibinfo{year}{1995}), \bibinfo{pages}{709--734}.
\newblock


\bibitem[\protect\citeauthoryear{Merritt, Heimbaugh, {LaChapell}, and
  Lee}{Merritt et~al\mbox{.}}{2013}]%
        {merritt_i_2013}
\bibfield{author}{\bibinfo{person}{Stephanie~M. Merritt},
  \bibinfo{person}{Heather Heimbaugh}, \bibinfo{person}{Jennifer {LaChapell}},
  {and} \bibinfo{person}{Deborah Lee}.} \bibinfo{year}{2013}\natexlab{}.
\newblock \showarticletitle{I Trust It, but I Don't Know Why: Effects of
  Implicit Attitudes Toward Automation on Trust in an Automated System}.
\newblock \bibinfo{journal}{\emph{Human Factors}} \bibinfo{volume}{55},
  \bibinfo{number}{3} (\bibinfo{year}{2013}), \bibinfo{pages}{520--534}.
\newblock


\bibitem[\protect\citeauthoryear{Nabi and Shpitser}{Nabi and Shpitser}{2018}]%
        {nabi2018fair}
\bibfield{author}{\bibinfo{person}{Razieh Nabi} {and} \bibinfo{person}{Ilya
  Shpitser}.} \bibinfo{year}{2018}\natexlab{}.
\newblock \showarticletitle{Fair inference on outcomes}. In
  \bibinfo{booktitle}{\emph{Proceedings of the... AAAI Conference on Artificial
  Intelligence. AAAI Conference on Artificial Intelligence}},
  Vol.~\bibinfo{volume}{2018}. NIH Public Access, \bibinfo{pages}{1931}.
\newblock


\bibitem[\protect\citeauthoryear{Narayanan}{Narayanan}{2018}]%
        {narayanan_translation_2018}
\bibfield{author}{\bibinfo{person}{Arvind Narayanan}.}
  \bibinfo{year}{2018}\natexlab{}.
\newblock \showarticletitle{Translation tutorial: 21 fairness definitions and
  their politics}. In \bibinfo{booktitle}{\emph{{ACM} Conference on Fairness,
  Accountability, and Transparency}}.
\newblock


\bibitem[\protect\citeauthoryear{Nikbin, Ismail, Marimuthu, and
  Abu-Jarad}{Nikbin et~al\mbox{.}}{2011}]%
        {nikbin_effects_2011}
\bibfield{author}{\bibinfo{person}{Davoud Nikbin}, \bibinfo{person}{Ishak
  Ismail}, \bibinfo{person}{Malliga Marimuthu}, {and} \bibinfo{person}{{Ismael
  Younis} Abu-Jarad}.} \bibinfo{year}{2011}\natexlab{}.
\newblock \showarticletitle{The effects of perceived service fairness on
  satisfaction, trust, and behavioural intentions}.
\newblock \bibinfo{journal}{\emph{Singapore Management Review}}
  \bibinfo{volume}{33}, \bibinfo{number}{2} (\bibinfo{year}{2011}),
  \bibinfo{pages}{58--73}.
\newblock


\bibitem[\protect\citeauthoryear{Ribeiro, Singh, and Guestrin}{Ribeiro
  et~al\mbox{.}}{2016}]%
        {ribeiro_why_2016}
\bibfield{author}{\bibinfo{person}{Marco~Tulio Ribeiro},
  \bibinfo{person}{Sameer Singh}, {and} \bibinfo{person}{Carlos Guestrin}.}
  \bibinfo{year}{2016}\natexlab{}.
\newblock \showarticletitle{"{Why} {Should} {I} {Trust} {You}?": {Explaining}
  the {Predictions} of {Any} {Classifier}}.
\newblock \bibinfo{journal}{\emph{arXiv:1602.04938 [cs, stat]}}
  (\bibinfo{date}{Feb.} \bibinfo{year}{2016}).
\newblock
\newblock
\shownote{arXiv: 1602.04938.}


\bibitem[\protect\citeauthoryear{Robert, Pierce, Marquis, Kim, and
  Alahmad}{Robert et~al\mbox{.}}{2020}]%
        {robert_designing_2020}
\bibfield{author}{\bibinfo{person}{Lionel~P. Robert}, \bibinfo{person}{Casey
  Pierce}, \bibinfo{person}{Liz Marquis}, \bibinfo{person}{Sangmi Kim}, {and}
  \bibinfo{person}{Rasha Alahmad}.} \bibinfo{year}{2020}\natexlab{}.
\newblock \showarticletitle{Designing fair AI for managing employees in
  organizations: a review, critique, and design agenda}.
\newblock \bibinfo{journal}{\emph{Human–Computer Interaction}}
  (\bibinfo{year}{2020}), \bibinfo{pages}{1--31}.
\newblock


\bibitem[\protect\citeauthoryear{Roy, Devlin, and Sekhon}{Roy
  et~al\mbox{.}}{2015}]%
        {roy_impact_2015}
\bibfield{author}{\bibinfo{person}{Sanjit~Kumar Roy}, \bibinfo{person}{James~F.
  Devlin}, {and} \bibinfo{person}{Harjit Sekhon}.}
  \bibinfo{year}{2015}\natexlab{}.
\newblock \showarticletitle{The impact of fairness on trustworthiness and trust
  in banking}.
\newblock \bibinfo{journal}{\emph{Journal of Marketing Management}}
  \bibinfo{volume}{31}, \bibinfo{number}{9-10} (\bibinfo{year}{2015}),
  \bibinfo{pages}{996--1017}.
\newblock


\bibitem[\protect\citeauthoryear{Srivastava, Heidari, and Krause}{Srivastava
  et~al\mbox{.}}{2019}]%
        {srivastava_mathematical_2019}
\bibfield{author}{\bibinfo{person}{Megha Srivastava}, \bibinfo{person}{Hoda
  Heidari}, {and} \bibinfo{person}{Andreas Krause}.}
  \bibinfo{year}{2019}\natexlab{}.
\newblock \showarticletitle{Mathematical Notions vs. Human Perception of
  Fairness: A Descriptive Approach to Fairness for Machine Learning}. In
  \bibinfo{booktitle}{\emph{Proceedings of the 25th ACM KDD}}.
  \bibinfo{pages}{2459–2468}.
\newblock


\bibitem[\protect\citeauthoryear{Symeonidis, Nanopoulos, and
  Manolopoulos}{Symeonidis et~al\mbox{.}}{2009}]%
        {Symeonidis_moviexplain_2009}
\bibfield{author}{\bibinfo{person}{Panagiotis Symeonidis},
  \bibinfo{person}{Alexandros Nanopoulos}, {and} \bibinfo{person}{Yannis
  Manolopoulos}.} \bibinfo{year}{2009}\natexlab{}.
\newblock \showarticletitle{MoviExplain: A Recommender System with
  Explanations}. In \bibinfo{booktitle}{\emph{Proceedings of the Third ACM
  Conference on Recommender Systems}}. \bibinfo{pages}{317--320}.
\newblock


\bibitem[\protect\citeauthoryear{van~den Bos}{van~den Bos}{2001}]%
        {van_den_bos_uncertainty_2001}
\bibfield{author}{\bibinfo{person}{Kees van~den Bos}.}
  \bibinfo{year}{2001}\natexlab{}.
\newblock \showarticletitle{Uncertainty management: The influence of
  uncertainty salience on reactions to perceived procedural fairness}.
\newblock \bibinfo{journal}{\emph{Journal of Personality and Social
  Psychology}} \bibinfo{volume}{80}, \bibinfo{number}{6}
  (\bibinfo{year}{2001}), \bibinfo{pages}{931--941}.
\newblock


\bibitem[\protect\citeauthoryear{Wang, Harper, and Zhu}{Wang
  et~al\mbox{.}}{2020}]%
        {wang_factors_2020}
\bibfield{author}{\bibinfo{person}{Ruotong Wang}, \bibinfo{person}{F.~Maxwell
  Harper}, {and} \bibinfo{person}{Haiyi Zhu}.} \bibinfo{year}{2020}\natexlab{}.
\newblock \showarticletitle{Factors Influencing Perceived Fairness in
  Algorithmic Decision-Making: Algorithm Outcomes, Development Procedures, and
  Individual Differences}. In \bibinfo{booktitle}{\emph{Proceedings of CHI
  2020}}. \bibinfo{pages}{1–14}.
\newblock


\bibitem[\protect\citeauthoryear{Yin, Vaughan, and Wallach}{Yin
  et~al\mbox{.}}{2018}]%
        {Yin_does_2018}
\bibfield{author}{\bibinfo{person}{Ming Yin}, \bibinfo{person}{Jennifer~Wortman
  Vaughan}, {and} \bibinfo{person}{Hanna Wallach}.}
  \bibinfo{year}{2018}\natexlab{}.
\newblock \showarticletitle{Does Stated Accuracy Affect Trust in Machine
  Learning Algorithms?}. In \bibinfo{booktitle}{\emph{Proceedings of ICML2018
  Workshop on Human Interpretability in Machine Learning ({WHI 2018})}}.
\newblock


\bibitem[\protect\citeauthoryear{Zhang, Liao, and Bellamy}{Zhang
  et~al\mbox{.}}{2020}]%
        {zhang_effect_2020}
\bibfield{author}{\bibinfo{person}{Yunfeng Zhang}, \bibinfo{person}{Q.~Vera
  Liao}, {and} \bibinfo{person}{Rachel K.~E. Bellamy}.}
  \bibinfo{year}{2020}\natexlab{}.
\newblock \showarticletitle{Effect of Confidence and Explanation on Accuracy
  and Trust Calibration in AI-Assisted Decision Making}. In
  \bibinfo{booktitle}{\emph{Proceedings of the 2020 Conference on Fairness,
  Accountability, and Transparency}} \emph{(\bibinfo{series}{FAT* ’20})}.
  \bibinfo{pages}{295–305}.
\newblock


\bibitem[\protect\citeauthoryear{Zhou, Bridon, Chen, Khawaji, and Wang}{Zhou
  et~al\mbox{.}}{2015a}]%
        {zhou_be_2015}
\bibfield{author}{\bibinfo{person}{Jianlong Zhou}, \bibinfo{person}{Constant
  Bridon}, \bibinfo{person}{Fang Chen}, \bibinfo{person}{Ahmad Khawaji}, {and}
  \bibinfo{person}{Yang Wang}.} \bibinfo{year}{2015}\natexlab{a}.
\newblock \showarticletitle{Be {Informed} and {Be} {Involved}: {Effects} of
  {Uncertainty} and {Correlation} on {User} {Confidence} in {Decision}
  {Making}}. In \bibinfo{booktitle}{\emph{Proceedings of {CHI}2015
  {Works}-in-{Progress}}}. \bibinfo{address}{Korea}.
\newblock


\bibitem[\protect\citeauthoryear{Zhou and Chen}{Zhou and Chen}{2018}]%
        {Zhou_human_2018}
\bibfield{editor}{\bibinfo{person}{Jianlong Zhou} {and} \bibinfo{person}{Fang
  Chen}} (Eds.). \bibinfo{year}{2018}\natexlab{}.
\newblock \bibinfo{booktitle}{\emph{Human and Machine Learning: Visible,
  Explainable, Trustworthy and Transparent}}.
\newblock \bibinfo{publisher}{Springer}, \bibinfo{address}{Cham}.
\newblock


\bibitem[\protect\citeauthoryear{Zhou, Hu, Li, Yu, and Chen}{Zhou
  et~al\mbox{.}}{2019}]%
        {zhou_physiological_2019}
\bibfield{author}{\bibinfo{person}{Jianlong Zhou}, \bibinfo{person}{Huaiwen
  Hu}, \bibinfo{person}{Zhidong Li}, \bibinfo{person}{Kun Yu}, {and}
  \bibinfo{person}{Fang Chen}.} \bibinfo{year}{2019}\natexlab{}.
\newblock \showarticletitle{Physiological Indicators for User Trust in Machine
  Learning with Influence Enhanced Fact-Checking}. In
  \bibinfo{booktitle}{\emph{Machine Learning and Knowledge Extraction}}.
  \bibinfo{pages}{94--113}.
\newblock


\bibitem[\protect\citeauthoryear{Zhou, Sun, Chen, Wang, Taib, Khawaji, and
  Li}{Zhou et~al\mbox{.}}{2015b}]%
        {zhou_measurable_2015}
\bibfield{author}{\bibinfo{person}{Jianlong Zhou}, \bibinfo{person}{Jinjun
  Sun}, \bibinfo{person}{Fang Chen}, \bibinfo{person}{Yang Wang},
  \bibinfo{person}{Ronnie Taib}, \bibinfo{person}{Ahmad Khawaji}, {and}
  \bibinfo{person}{Zhidong Li}.} \bibinfo{year}{2015}\natexlab{b}.
\newblock \showarticletitle{Measurable {Decision} {Making} with {GSR} and
  {Pupillary} {Analysis} for {Intelligent} {User} {Interface}}.
\newblock \bibinfo{journal}{\emph{ACM Transactions on Computer-Human
  Interaction}} \bibinfo{volume}{21}, \bibinfo{number}{6}
  (\bibinfo{year}{2015}), \bibinfo{pages}{33}.
\newblock


\end{thebibliography}

\end{document}